# THE NUMERICAL SIMULATION OF A 3D FLOW IN THE VKI-GENOA TURBINE CASCADE TAKING INTO ACCOUNT THE LAMINAR-TURBULENT TRANSITION

SERGIY YERSHOV [1], ANTON DEREVYANKO [1],
VIKTOR YAKOVLEV [1] AND MARIA GRYZUN [2]

[1] *Institute for Mechanical Engineering Problems of National Academy of Sciences, 2/10 Pozharsky St., 61046 Kharkiv, Ukraine*
[2] *National Technical University "Kharkiv Polytechnic Institute" 21 Frunze St., 61002 Kharkiv, Ukraine*

**Abstract**: This study presents a numerical simulation of a 3D viscous flow in the VKI-Genoa cascade with taking into account the laminar-turbulent transition. The numerical simulation is performed using the Reynolds-averaged Navier-Stokes equations and the two-equation *k*-ω SST turbulence model. The algebraic Production Term Modification model is used for modeling the laminar-turbulent transition. Computations of both fully turbulent and transitional flows are carried out. The contours of the Mach number, the turbulence kinetic energy, the entropy function, as well as limiting streamlines are presented. Our numerical results demonstrate the influence of the laminar-turbulent transition on the secondary flow pattern. The comparison between the present computational results and the existing experimental and numerical data shows that the proposed approach reflects sufficiently the physics of the laminar-turbulent transition in turbine cascades.

## 1. Introduction

Despite the fact that turbomachines have been known for a long time and a large number of scientific papers have been devoted to their research and improvement, not all possibilities of such improvement have been exhausted yet. The application of modern methods for flow computations using 3D Reynolds-averaged Navier-Stokes (RANS) solvers [1] permits the numerical flow field simulation in a turbomachinery and the 3D design of their flowpathes. However, if such design does not take into account all phenomena of a 3D turbulent viscous compressible flow, then an incorrect estimation of the turbomachinery efficiency is practically inevitable, and therefore there remains room for its improvement.

The laminar-turbulent transition of a turbomachinery flow is one of such insufficiently studied phenomena, and therefore it is usually unaccounted in most studies. The effect of transition on both the kinetic energy losses and the turbomachinery efficiency is ambiguous. As it is known, the kinetic energy losses are smaller in the laminar boundary layer when compared with that in the turbulent boundary layer of the same thickness, but the former is more susceptible to separation, in which the losses can increase [2]. On the other hand, the flow acceleration can cause a relaminarization and thinning of the boundary layer. Thinner boundary layers are more resistant to separation, and even if a separation occurs, then its thickness will be smaller than that of a thicker turbulent boundary layer in slightly-accelerated flow cascades. In this case, again, the kinetic energy losses may be less when compared with the fully turbulent flow case, and this fact is often used to design highly-loaded cascades. Both the effect of the 3D flow pattern on laminar-turbulent transition and the reverse effect of laminar-turbulent transition on 3D secondary flows are practically unstudied. Such ambiguity and uncertainty of the influence of laminar-turbulent transitions on the turbomachinery efficiency requires a complete 3D flow investigation for every flowpath design to evaluate its respective performance.

There exist several models that describe laminar-turbulent transitions and could be used with the RANS equations. Sufficiently detailed reviews, concerning transition models, are presented in [3-5]. Such models are mainly semi-empirical and are not universal, so each model has a range of applicability. Therefore, the development of new transition models and their verification with respect to a variety of turbulence models is an important problem.

This study considers a numerical simulation of a 3D transitional flow in a turbine cascade. A simple algebraic model is chosen as a laminar-to-turbulent transition model. It requires less computational resources in comparison with approaches that introduce additional differential equations [6]. The numerical solutions for two flow cases, namely a fully turbulent flow and a transitional flow, are compared with each other as well as with the existing experimental and numerical data.

## 2. Mathematical model of the flow

A 3D viscous compressible flow through a turbine cascade is described by the RANS equations, written in a local curvilinear coordinate system that rotates at a constant speed Ω:

$$\frac{\partial QJ}{\partial t}+\frac{\partial F^j}{\partial \xi^j}=H,$$

where $Q$ is the vector of conservative variables in the Cartesian coordinates; $F^j=F_i\xi_i^jJ$ is the flux vector in the curvilinear coordinates; $F_i$ is the flux vector in the Cartesian coordinates; $t$ is time; $\xi^j$ are the curvilinear coordinates; $\xi_i^j$ are the metric coefficients; $H$ is the source term vector; $J$ is the Jacobian of the coordinate transformation between the Cartesian and curvilinear coordinate systems.

If the rotation occurs around the $x_3$-axis, then the vector of conservative variables $Q$, the flux vector $F_i$ and the source term vector $H$ in the Cartesian coordinate system $x_i$ are of the form:

$$Q=\begin{pmatrix}\rho\\ \rho u_1\\ \rho u_2\\ \rho u_3\\ \rho h\end{pmatrix},\quad F_i=\begin{pmatrix}\rho u_i\\ \rho u_i u_1+\delta_{i1}p-\tau_{i1}\\ \rho u_i u_2+\delta_{i2}p-\tau_{i2}\\ \rho u_i u_3+\delta_{i3}p-\tau_{i3}\\ (\rho h+p)u_i-\tau_{ik}u_k+q_i\end{pmatrix},\quad H=J\begin{pmatrix}0\\ 2\rho u_2\Omega+\rho\Omega^2 r_1\\ -2\rho u_1\Omega+\rho\Omega^2 r_2\\ 0\\ 0\end{pmatrix},$$

where $\rho$, $u_i$, and $p$ are the density, the components of the velocity vector, and the pressure, respectively; $\tau_{ik}$ is the effective stress tensor; $q_i$ is the effective heat flux; $\gamma$ is the specific heat ratio; $r$, $r_1$, and $r_2$ are the distance from the rotation axis and its projections on the coordinate axes $x_1$ and $x_2$, respectively; $h=\frac{1}{\gamma-1}\frac{p}{\rho}+\frac{u_iu_i-\Omega^2r^2}{2}+k$; $k$ is the turbulence kinetic energy (TKE); $\delta_{ij}$ is the Kronecker delta.

The metric coefficients and the Jacobian of the coordinate transformation are written as

$$\xi_i^j=\frac{\partial \xi^j}{\partial x_i}=\lim_{\Delta x_i\to 0}\frac{\Delta\xi^j}{\Delta x_i},\quad J=\left|\xi_i^j\right|=\begin{vmatrix}\xi_1^1 & \xi_1^2 & \xi_1^3\\ \xi_2^1 & \xi_2^2 & \xi_2^3\\ \xi_3^1 & \xi_3^2 & \xi_3^3\end{vmatrix}=\varepsilon_{jkl}\xi_1^j\xi_2^k\xi_3^l,$$

where $\varepsilon_{jkl}$ is the Levi-Civita symbol.

The effective stress tensor represents a sum of the viscous and Reynolds (turbulent) stress tensors:

$$\tau_{ik}=\bar{\tau}_{ik}+\hat{\tau}_{ik},$$

where $\bar{\tau}_{ik} = 2\mu\left(S_{ik} - \frac{1}{3}S_{ll}\delta_{ik}\right)$ is the viscous stress tensor; $S_{ik} = \frac{1}{2}\left(\frac{\partial u_i}{\partial x_j} + \frac{\partial u_j}{\partial x_i}\right)$ is the strain rate tensor; $\mu = \rho\nu$ is the dynamic viscosity; $\nu$ is the kinematic viscosity.

Similarly, the effective heat flux is a sum of the molecular and turbulent heat fluxes:

$$q_i = \bar{q}_i + \hat{q}_i,$$

where $\bar{q}_i = \bar{\lambda}\frac{\partial T}{\partial x_i}$ is the molecular heat flux; $\hat{q}_i$ is the turbulent heat flux; $\bar{\lambda}$ is the molecular thermal conductivity; $T$ is the temperature.

Here and further, the derivatives in the Cartesian coordinates are determined through the curvilinear coordinate derivatives and vice versa according to the following coordinate transformations:

$$\frac{\partial}{\partial x_i} = \frac{\partial}{\partial \xi^j}\frac{\partial \xi^j}{\partial x_i} = \xi_i^j \frac{\partial}{\partial \xi^j}, \quad \frac{\partial}{\partial \xi^j} = \frac{\partial}{\partial x_i}\frac{\partial x_i}{\partial \xi^j} = x_i^j \frac{\partial}{\partial x_i}.$$

**3. Turbulence modeling**

For modeling turbulence in the present study, we use the two-equation $k$-$\omega$ SST model developed by Menter [7]. The model is written in the low-Reynolds form [8] that takes into account the Production Term Modification (PTM) [9-11]. The turbulence and flow equations in the resulting system are solved separately, which enables more efficient computational algorithms.

The differential equations of the model are written in the form:

$$\frac{\partial VJ}{\partial t} + \frac{\partial W^j}{\partial \xi^j} = P - D - L,$$

where $V$ is the vector of the conservative turbulent variables in the Cartesian coordinates; $W^j = W_i\xi_i^j J$ is the turbulent flux vector in the curvilinear coordinates; $W_i$ is the turbulent flux vector in the Cartesian coordinates; $P, D$, and $L$ are the source term vectors. These vectors can be written as

$$V = \begin{pmatrix}\rho k\\ \rho\omega\end{pmatrix},\ W_i = \begin{pmatrix}\rho k u_i - (\bar{\mu} + \sigma_k\hat{\mu})\frac{\partial k}{\partial x_i}\\ \rho\omega u_i - (\bar{\mu} + \sigma_\omega\hat{\mu})\frac{\partial \omega}{\partial x_i}\end{pmatrix},\ P = \begin{pmatrix}P_k\\ P_\omega\end{pmatrix},\ D = \begin{pmatrix}D_k\\ D_\omega\end{pmatrix},\ D = \begin{pmatrix}D_k\\ D_\omega\end{pmatrix},\ L = \begin{pmatrix}0\\ L_\omega\end{pmatrix},$$

where $\omega$ is the specific dissipation rate; $P_k$ and $P_\omega$ are production of turbulence and production of dissipation, respectively; $D_k$ and $D_\omega$ are the dissipation of turbulence and the dissipation of dissipation, respectively; $L_\omega$ is cross-diffusion.

The turbulence production is written as follows

$$P_k = \alpha_{PTM}\hat{\tau}_{ij}S_{ij},$$

where $\hat{\tau}_{ij} = 2\hat{\mu}\left(S_{ij} - \frac{1}{3}S_{nn}\delta_{ij}\right) - \frac{2}{3}\rho k\delta_{ij}$ is the Reynolds stress tensor; $\alpha_{PTM}$ is the production term modifier [10], which equals 1 in the case of the standard high- and low-Reynolds models.

The dissipation production could be written in the following form:

$$P_\omega = \frac{\alpha\rho}{\hat{\mu}}\hat{\tau}_{ij}S_{ij}.$$

The dissipation of turbulence and the dissipation of dissipation could be defined using the turbulence variables, $k$ and $\omega$:

$$D_k = \beta^*\rho\omega k,\ D_\omega = \beta\rho\omega^2.$$

The cross-diffusion term, which is usually not taken into account in most of the $k-\varepsilon$-type turbulence models, is represented as follows

$$L_\omega = (1-F_1)\rho\sigma_{\omega 2}\frac{1}{\omega}\frac{\partial k}{\partial x_j}\frac{\partial \omega}{\partial x_j},$$

where $F_1$ is the first Menter's blending function, which depends on the distance from the wall, $y$, and is calculated using the following formulas:

$$F_1 = \max\left[\tanh\left(\arg_1^4\right), F_4\right],\ \arg_1 = \min\left[\max\left(\frac{\sqrt{k}}{\beta^*\omega y}, \frac{500\nu}{\omega y^2}\right), \frac{4\rho\sigma_{\omega 2}k}{CD_{k\omega}y^2}\right],$$

$$CD_{k\omega} = \max\left(2\rho\sigma_{\omega 2}\frac{1}{\omega}\frac{\partial k}{\partial x_j}\frac{\partial \omega}{\partial x_j}; 1{,}0\times 10^{-20}\right),\ F_4 = \exp\left[-\left(\frac{R_y}{120}\right)^8\right],\ R_y = \frac{y\sqrt{k}}{\nu}.$$

The correction function, $F_4$, which depends on the local turbulent Reynolds number, $R_y$, is introduced as a necessary modification for modeling the laminar-turbulent transition and should be equal to zero when the standard high- and low Reynolds models are used.

The dynamic turbulent viscosity is calculated using the turbulence parameters, $k$ and $\omega$, taking into account the Bradshaw's hypothesis [7] and the realizability constraints [12]:

$$\hat{\mu} = \alpha^*\frac{\rho k}{\max\left(\omega; \frac{\alpha^* F_2 S}{a_1}; \frac{\sqrt{3}}{2}\sqrt{S^2 - \frac{2}{3}S_{nn}^2}\right)},$$

where $S = \sqrt{2S_{ij}S_{ij}}$ is the magnitude of the mean strain rate tensor. The second Menter's blending function $F_2$ is computed as follows

$$F_2 = \tanh\left(\arg_2^2\right),\ \arg_2 = \min\left[\frac{2\sqrt{k}}{\beta^*\omega y}, \frac{500\nu}{\omega y^2}\right].$$

Values $\alpha^*, \alpha, \beta^*, \beta, \sigma_k, \sigma_\omega$ are found using a linear interpolation:

$$\varphi = F_1\varphi_1 + (1-F_1)\varphi_2,\ \varphi = \left|\alpha^*, \alpha, \beta^*, \beta, \sigma_k, \sigma_\omega\right|^T,$$

where the parameters, $\varphi_1$ and $\varphi_2$, correspond to the $k$–$\omega$ and $k-\varepsilon$ models, respectively; Menter's blending function $F_1$ is the interpolation weight coefficient here.

In the standard high-Reynolds Menter's $k$–$\omega$ SST model, the values $\varphi_1$ and $\varphi_2$ are constants. In the low-Reynolds model, some of these quantities that are related to the $k$–$\omega$ model and determine the behaviour of the boundary layer turbulence are defined using the local turbulent Reynolds number, $R_t = k/(\nu\omega)$:

$$\alpha_1^* = \frac{1/40 + R_t/6}{1 + R_t/6},\ \alpha_1 = \frac{5}{9}\cdot\frac{1/10 + R_t/2{,}7}{1 + R_t/2{,}7},\ \beta_1^* = 0.09\cdot\frac{5/18 + (R_t/8)^4}{1 + (R_t/8)^4},\ \beta_1 = 0.075,$$

$$\sigma_{k1} = 0.85,\ \sigma_{\omega 1} = 0.5,$$

$$\alpha_2^* = 1,\ \alpha_2 = 0.4403,\ \beta_2^* = 0.09,\ \beta_2 = 0.0828,\ \sigma_{k2} = 1.0,\ \sigma_{\omega 2} = 0.856,\ a_1 = 0.31.$$

**4. The laminar-turbulent transition model**

The basic idea behind the PTM transition model is in the reduction of the turbulence production in the laminar and transitional boundary layers by a factor of $\alpha_{PTM}$, which can be considered as an analogue of the turbulence intermittency coefficient.

It is assumed that the laminar-turbulent transition is affected by the turbulence intensity of an external flow and the local pressure gradient. Thus, the boundary layer turbulization is reproduced due to the impact of a high external flow turbulence intensity and of separation or preseparation flow conditions with an adverse pressure gradient.

The function $P_{tm1}$ accounts for the influence of the external flow turbulence intensity, and it is calculated as follows

$$P_{tm1} = 1 - c_{\mathrm{PTM}} \begin{cases} \left[\left(3.328\cdot 10^{-4}\right)R_\nu - \left(3.94\cdot 10^{-7}\right)R_\nu^2 + \left(1.43\cdot 10^{-10}\right)R_\nu^3\right], & R_\nu < 1000 \\ \left[0.12 + \left(1{,}0\cdot 10^{-5}\right)R_\nu\right], & R_\nu \geq 1000 \end{cases},$$

where $R_\nu = \dfrac{y^2 S}{\nu}$ is the Reynolds number, determined by the distance from the nearest wall and the magnitude of the mean strain rate tensor; $c_{\mathrm{PTM}}$ is a constant.

The function $P_{tm2}$ estimates the influence of the pressure gradient:

$$P_{tm2} = \begin{cases} -|K|^{0.4}\dfrac{R_\nu}{80}, & K < 0 \\ 0, & K \geq 0 \end{cases}, \quad K = -\frac{\mu}{\rho^2 U^3}\left(1 - M^2\right)\frac{dp}{ds},$$

where $U$ is the flow velocity; $M$ is the Mach number; $s$ is the streamwise coordinate.

The combined effect of the two factors is taken into account using the following relation:

$$P_{tm} = 1 - 0.94\left(P_{tm1} + P_{tm2}\right)F_3 \,\mathrm{th}\left[(y^+/17)^2\right],$$

where $F_3 = \exp\left[-(R_t/a_{\mathrm{PTM}})^{b_{\mathrm{PTM}}}\right]\left[1 - P(R_t)\right] - 1/2P(R_t)$ is a function that triggers the TKE production when the local Reynolds number reaches its critical value; $y^+ = yu_\tau/\nu$ is the dimensionless distance from the nearest wall; $u_\tau = \sqrt{\tau_w/\rho_w}$ is the friction velocity; $P(R_t) = \dfrac{2{,}5}{\sqrt{2\pi}}\exp\left[-\dfrac{(R_t - 3)^2}{2}\right]$ is a function of the turbulent Reynolds number $R_t$; $a_{\mathrm{PTM}}$ and $b_{\mathrm{PTM}}$ are constants.

Usually, the values of the constants are the following:

$a_{\mathrm{PTM}} = 3.45$; $b_{\mathrm{PTM}} = 2.0$; $c_{\mathrm{PTM}} = 1.0$.

According to the physics of transitional flows, the production term modifier $\alpha_{\mathrm{PTM}}$, similar to the turbulence intermittency, should be limited between zero and one. Therefore, in this study, the following restriction is imposed:

$$\alpha_{\mathrm{PTM}} = \min\left(1, P_{tm}\right).$$

## 5. Numerical approach

The governing equations are integrated numerically using an implicit second-order accurate ENO-scheme [13] that applies an exact Riemann solver to calculate fluxes at discretization cell boundaries. A simplified multigrid method, local time stepping, and Newton iterations [14] are used to accelerate the convergence of the proposed scheme. When performing calculations using a large Courant number, the time step should be adjusted for excessively elongated cells.

The considered approach has been implemented in the CFD solver F [15, 16].

## 6. Methodology of carrying out the calculations

In this study we found that the accuracy and the reliability of numerical results depend on several factors. First, in order to model adequately the laminar-turbulent transition, it is necessary to use physically grounded turbulence models. In particular, an important point is the use of the realizability constraints for the Reynolds stress tensor, without which the solution can be infeasible.

Second, in order to successfully model transitional behaviour, the number of cells and the discretization mesh quality must be severely restricted. In the transition zone of the boundary layer flow characteristics are subjected to rapid changes: the boundary layer thickness and its profile vary significantly within very short distances. Therefore, it is necessary to ensure that the mesh resolution is high along the streamwise and wall-normal directions. According to author's experience in computing transitional flows, an acceptable description of the laminar-turbulent transition in a three-dimensional cascade requires ensuring the following conditions:

- $y^+$ value for the cell nearest to the wall must be of the order of 1;
- there must be 30 or more cells across the boundary layer in the transition zone;
- there must be at least 150 cells along each blade surface in the stream-wise direction;
- cell sizes must change smoothly.

These requirements are usually realizable for computational meshes of size greater than between a few million and a few tens of million cells inside one 3D blade-to-blade passage.

Third, the convergence rate in the case of the transitional flows is quite slow, and it is recommended to use a fully turbulent flow field as an initial approximation. Such computations require a significant of running time, and it is highly desirable to apply the numerical methods that are able to operate on very fine meshes using the Courant number that is much greater than 1.

## 7. Computational results

We consider the three-dimensional flow in the cascade VKI-Genoa, which has been previously studied experimentally [17] and computationally [6, 18]. We use a computational mesh of H-type with an approximate orthogonalization of mesh lines near solid walls. The considered mesh contains on the order of 4.2 million cells (128x128x256). The value of $y^+$ of the cell nearest to the wall is approximately 1; more than 30 cells are located in the transition region across the boundary layer, and there are 168 cells along the blade surface in the streamwise direction with refinement at the leading and trailing edges.

The flow through the considered cascade is subsonic with the exit Mach number $M_{2is} = 0.24$ and the Reynolds number $\mathrm{Re} = 1.6 \cdot 10^6$. The turbulence intensity at the inlet equals 1 percent. The experiment did not determine the endwall boundary layer thickness in front of the cascade. Since an assignment of this value is mandatory for a three-dimensional cascade flow calculation, in the present study we arbitrarily assume that the boundary layer originates at the inlet boundary of the computational domain, which was 1 axial chord upstream of the leading edges.

Shown in Figure 1 is the distribution of the dimensionless adiabatic speed $U$ at the blade surface in the mid-span section. The nondimensionalization is performed by dividing by the inlet velocity $U_0$. The surface coordinate $S$ is measured along the blade contour from the leading edge. The value $S_{\max}$ corresponds to the surface length from the leading edge to the trailing edge. The dots display the experimental data [17], the black lines present the numerical results of Langtry [6], the red and blue lines demonstrate the numerical results for the fully turbulent and transitional flow cases obtained in the present study, respectively. There is a good agreement between these results for the most part of the blade surface, except for the trailing edge region.

Displayed in Figure 2 are the Mach number contours in the mid-span section of the cascade, which are calculated in fully turbulent and transitional flow cases. The behaviour of main flow is similar enough in both cases, and flow structures appear to be qualitatively and quantitatively very close.

Figure 3 presents the distribution of the dimensionless friction velocity, $u_\tau = \sqrt{\tau_w/\rho}$, along the blade surface in the mid-span section. Here $\tau_w$ is the wall shear stress. The nondimensionalization is performed by dividing by the local adiabatic speed. The dots show the experimental data [17], the black line displays the numerical results of Langtry [6], the red and blue lines demonstrate the results for the fully turbulent and transitional flow cases obtained in the present study, respectively. We took the results of other authors, shown in the Figure, from [6]. Figure 4 gives the analogous graph, but the data of other authors are taken from [18]. As can be seen, the interpretations of the experimental results in [6] and [18] are somewhat different. Nevertheless, the following points can be concluded:

- the transition occurs in the non-separated boundary layer at the blade suction side of the VKI-Genoa cascade;
- the model of the laminar-turbulent transition that is used in the present study reflects qualitatively correctly the transition phenomenon;

- the transition model predicts the transition point somewhat upstream, compared with the model [6], and somewhat downstream, compared with the model [18]; it should be noted that the differential models of [6] and [18] differ mainly in the empirical calibration of the correlation functions;

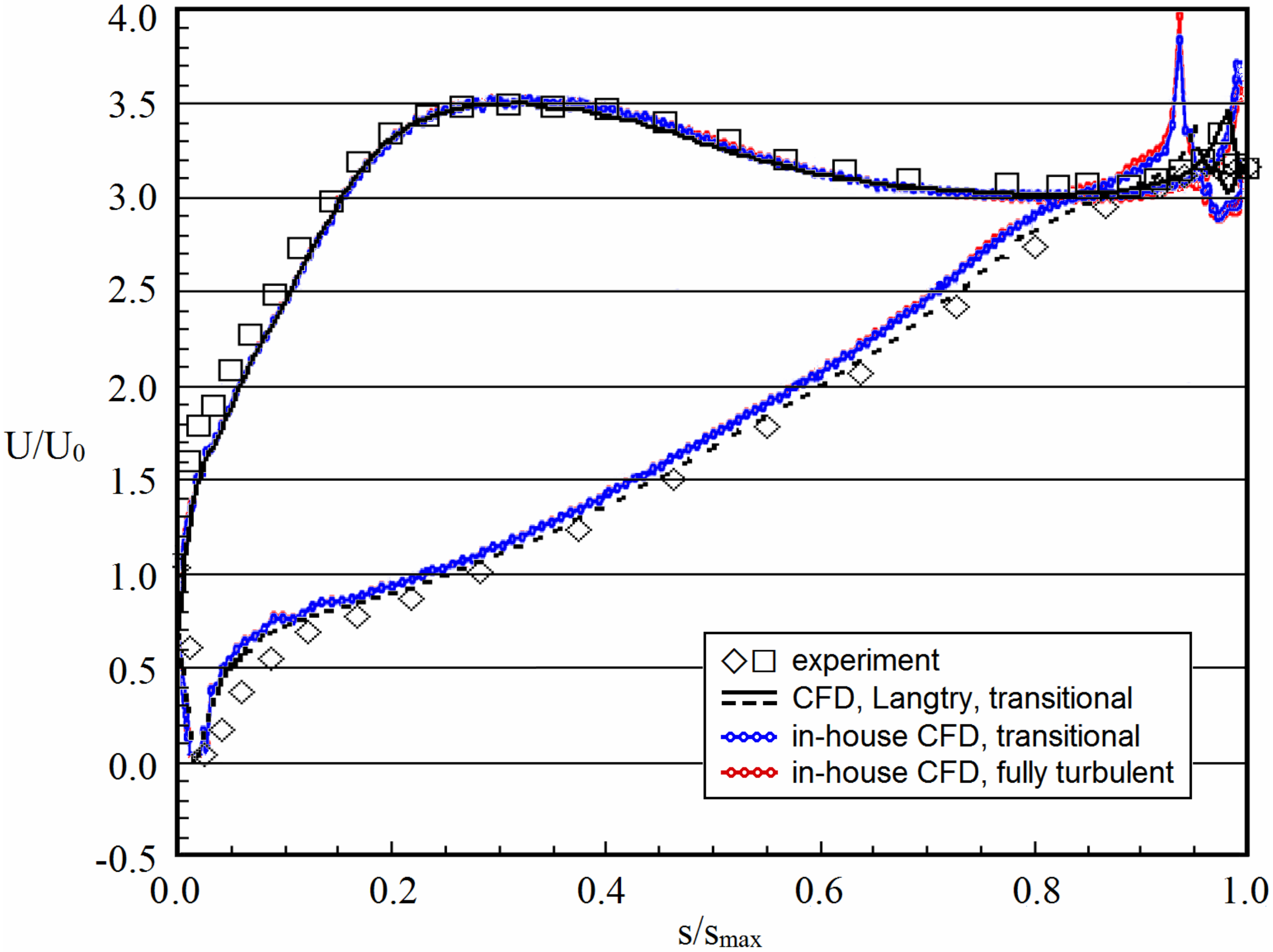

Fig. 1. The dimensionless adiabatic velocity at the blade surface in the mid-span section

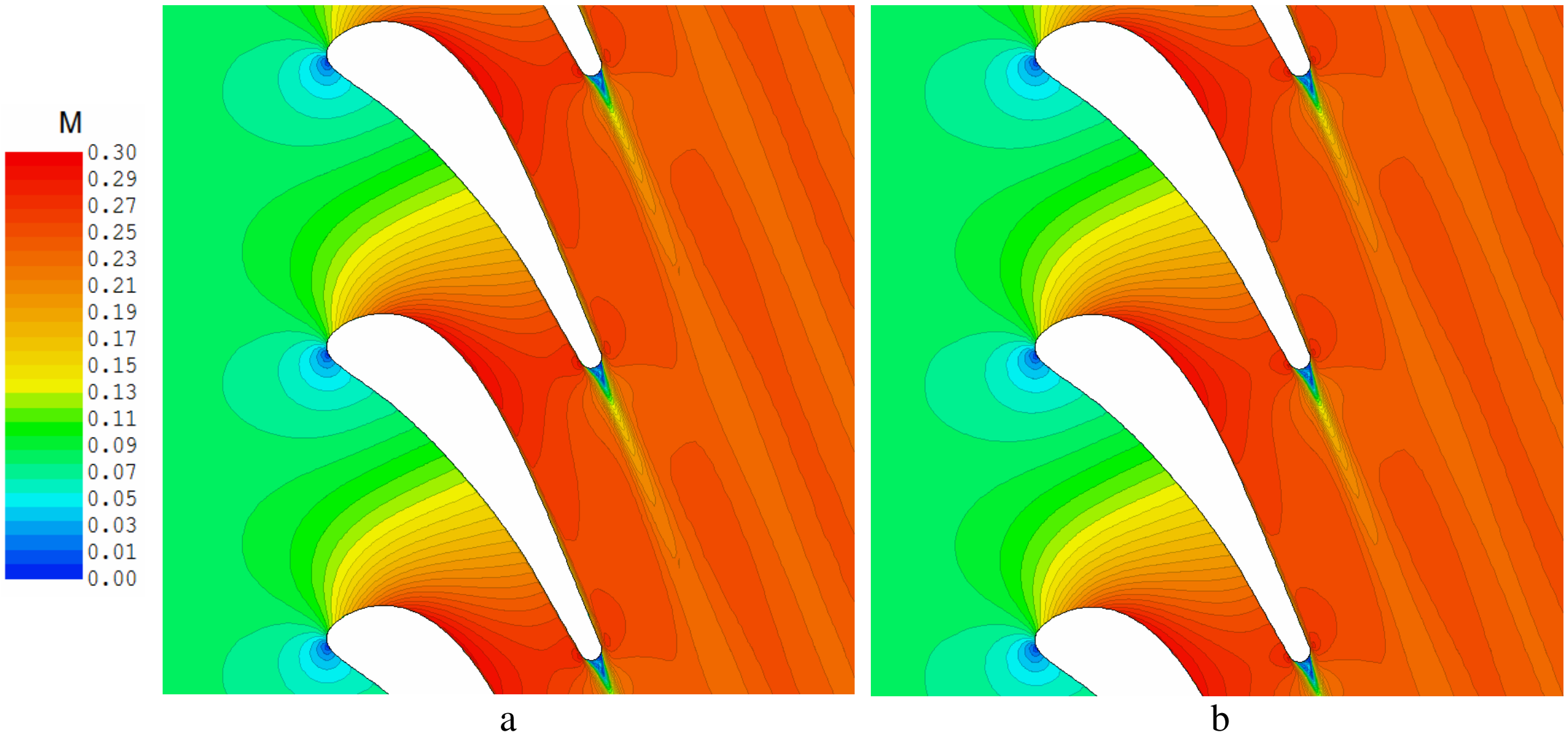

Fig. 2. The Mach number contours in the mid-span section
a – fully turbulent flow; b – transitional flow

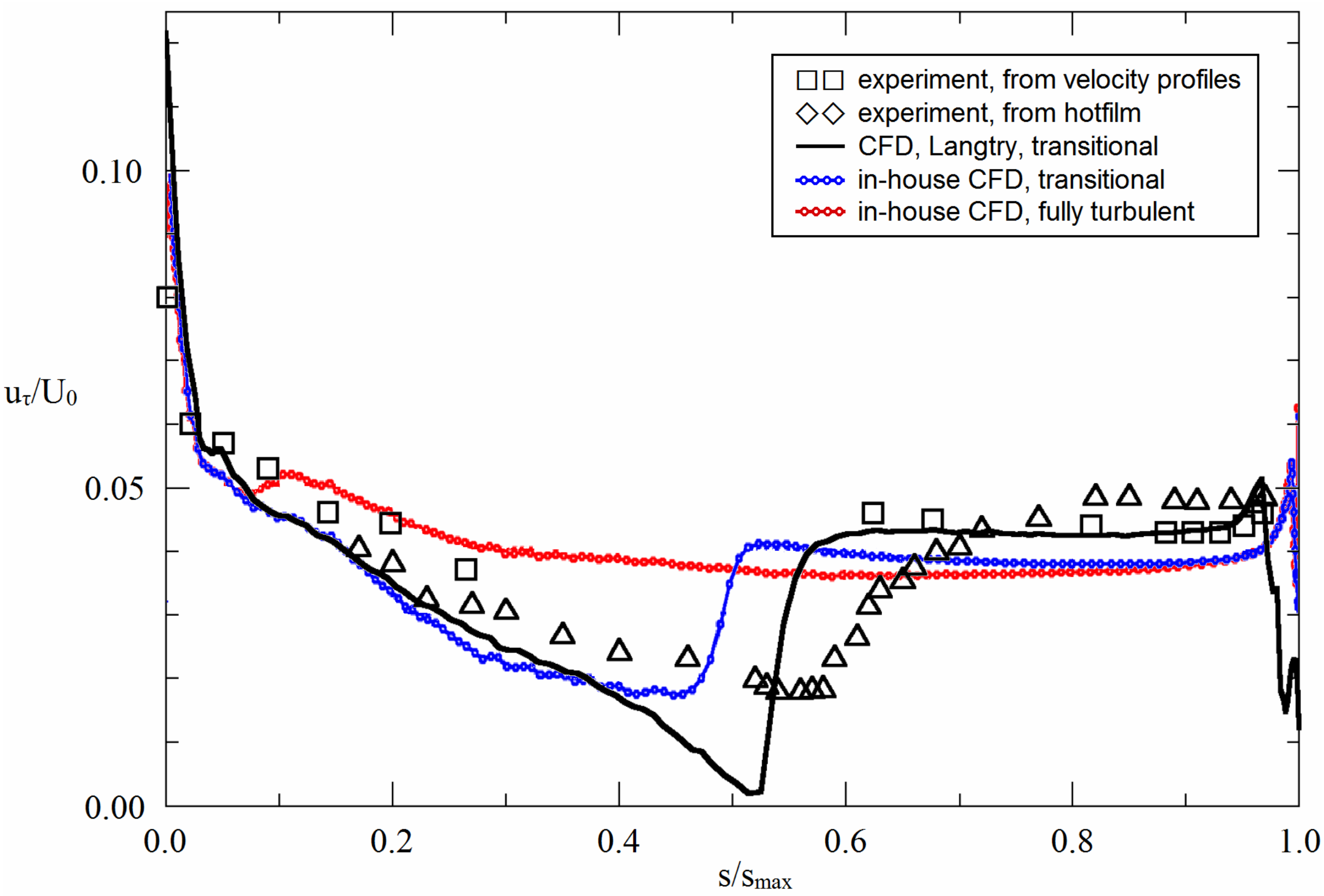

Fig. 3. The dimensionless friction velocity at the blade suction surface in the mid-span section. Based on data from [6]

- the transition modeling accuracy can be considered to be quite acceptable, since in the experiment [17] the transition point locates at $S/S_{\max}$=0.48, which corresponds approximately to the middle of the transition region predicted in this study;
- the peak position misalignment near the trailing edge, as well as that observed in Figure 1, is apparently due to the different definition of the surface length $S_{\max}$, but in any case the error does not exceed in order the trailing edge thickness.

It should be noted that the preliminary 2D flow calculations have shown a later transition along with slightly larger both the exit flow velocity and the Reynolds number.

Displayed in Figure 5 are the TKE contours for the cascade mid-span section, which are obtained in simulating the fully turbulent and transitional flow cases. It is seen that in the case of the fully turbulent flow (Figure 5, a) the TKE growth at the blade suction surface begins almost immediately downstream of the leading edge, but at the pressure side it starts to grow somewhat later. When taking into account the transition (Figure 5, b), the TKE growth occurs approximately at the cascade throat: at the trailing edge of the pressure side and near the middle of the chord at the suction side. Thus, the flow upstream of the cascade throat remains rather laminar. It should be noted that in the case of the transitional flow the maximum TKE values downstream of the transition point are higher and the turbulent boundary layer thickness is somewhat less, than those of the fully turbulent flow case.

Demonstrated in Figure 6 is the limiting streamlines on the endwall surface. Here and below streamline tracing is performed using the freeware, Paraview [19]. The flow patterns of the fully turbulent and transitional flow cases seem to be qualitatively similar. The main difference is that the pressure-driven cross-flow in the endwall boundary layer from the blade pressure side to the suction side is more intensive in the case of the transitional flow, especially in the area between the branches of the horseshoe vortices. In this connection, the horseshoe vortex branch that runs along the suction side flows on the blade suction side somewhat upstream in the case of the transitional flow when compared with the fully turbulent flow case. The positions of the second branch of the horseshoe vortex are about the same in both flow cases.

We may assume that the endwall boundary layer in the region of the horseshoe vortex is rather laminar. The profile of such the boundary layer is less filled than that of the turbulent boundary layer, and therefore, it is more susceptible to both separation and cross-flow. The turbulent flow develops downstream of the horseshoe vortex. The profile of the boundary layer in this area becomes more filled. Therefore, in the considered fully turbulent and transitional flow cases, the difference in the intensity of the endwall cross-flow in the region between the blade pressure side and the nearest horseshoe vortex branch is less pronounced.

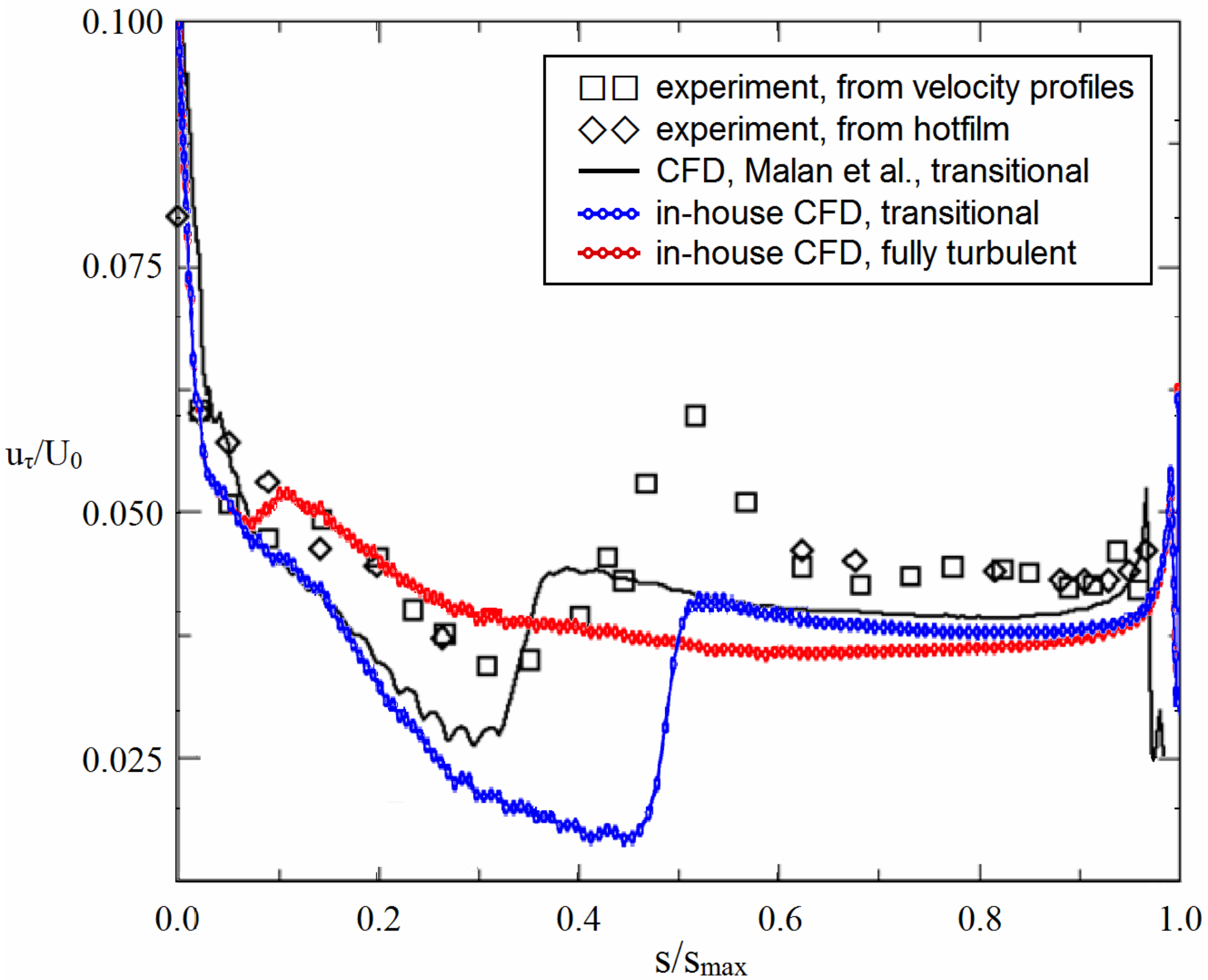

Fig. 4. The dimensionless friction velocity at the blade suction surface in the mid-span section
Based on data from [18]

Some confirmation of this assumption can be found in Figure 7, which gives the TKE contours in the endwall boundary layer at the distance of 1 percent of the blade span from the endwall. Since the endwall boundary layer starts to emerge from the inlet boundary of the computational domain, it remains sufficiently thin and quasi-laminar in the blade-to-blade passage even when modeling a fully turbulent flow. It is seen that the flow turbulization (TKE increasing) begins in the horseshoe vortex upstream of the leading edge. The values of TKE in this area are relatively small and are about the same for both the turbulent and transitional flow cases. Downstream, in these flow cases there is a significant increase of TKE in the region of the boundary layer cross-flow. This process in the case of the transitional flow starts upstream when compared with the fully turbulent flow case, and approximately in the same zone where the boundary layer is turbulized at the blade suction surface.

Figure 8 shows the TKE contours at the distance of 0.1 percent of the cascade pitch from the blade suction surface in both flow cases. It is clearly seen that, as mentioned earlier, the growth of TKE in the case of the fully turbulent flow begins immediately downstream of the leading edges, whereas in the transitional flow case it takes place near the cascade throat. In both flow cases, the turbulization of the blade suction side boundary layer occurs almost simultaneously over a whole blade span. The observed zones of the turbulization delay and advance in the secondary flows region correlate well with the position of the boundary layer cross-flow at the blade suction side and with the related areas of high and low flow velocity.

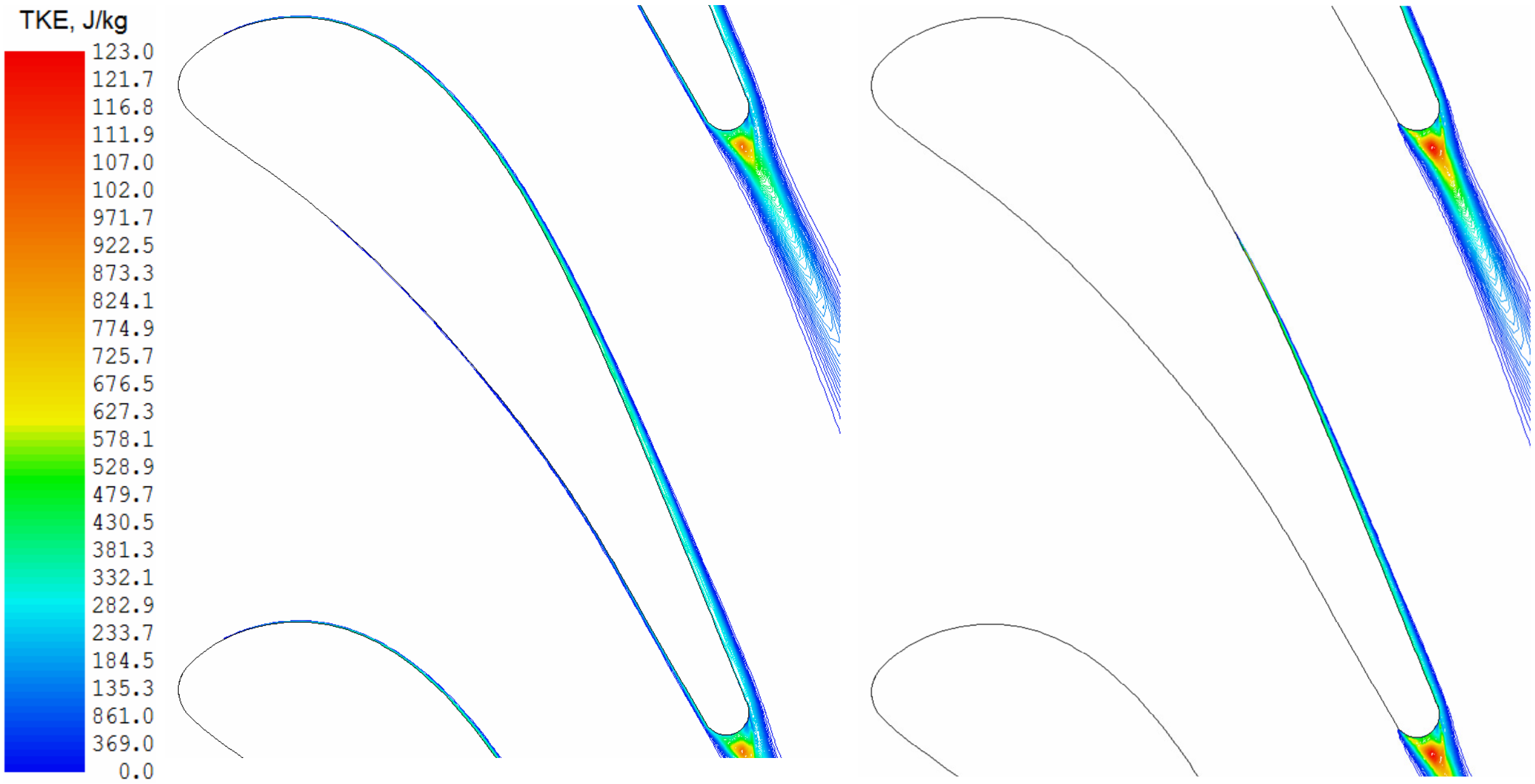

a b

Fig. 5. The TKE contours in the mid-span section
a – fully turbulent flow; b – transitional flow

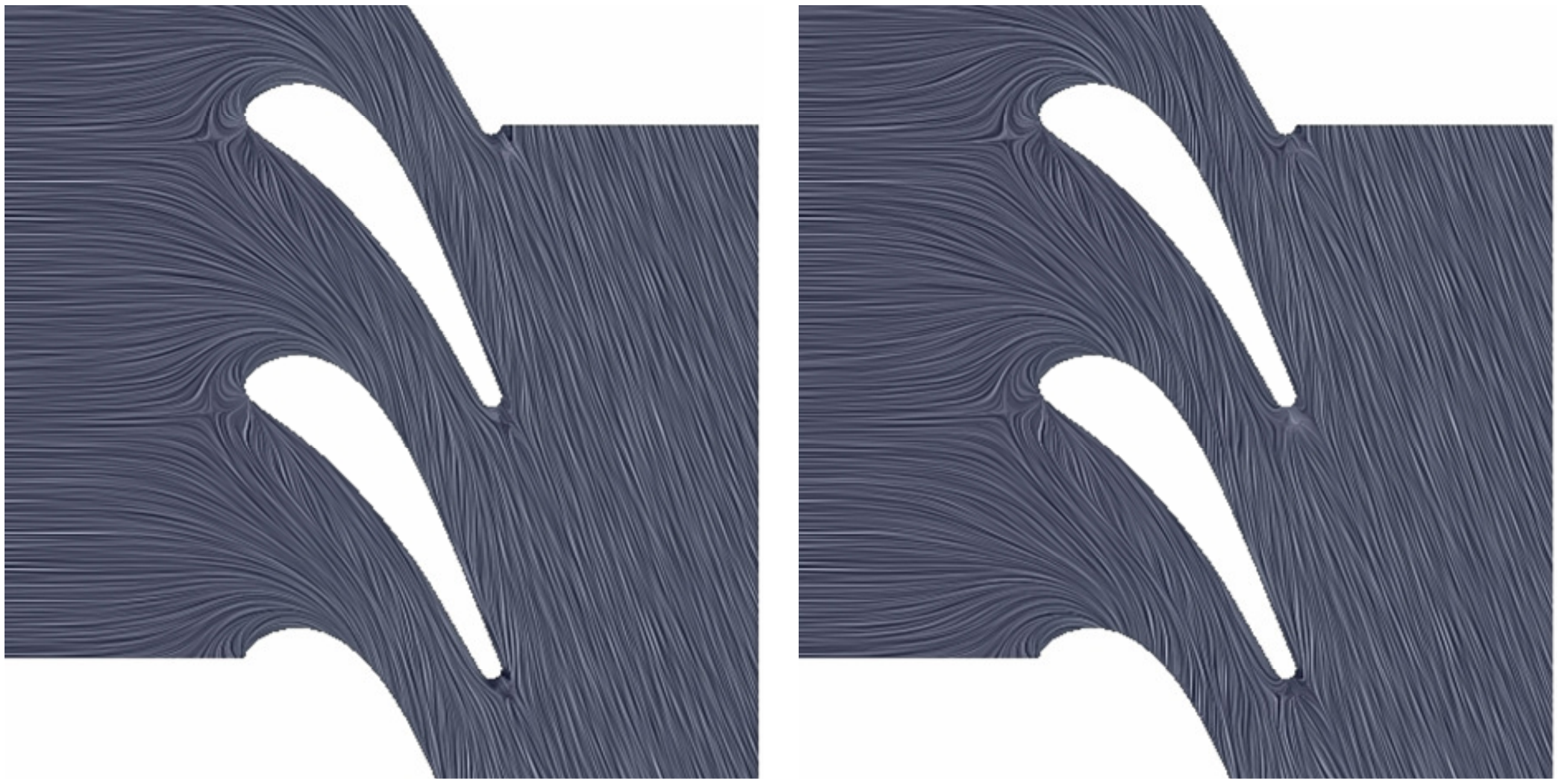

a b

Fig. 6. The limiting streamlines at the endwall surface
a – fully turbulent flow; b – transitional flow

Given in Figure 9 are limiting streamlines at the trailing edge and in the downstream endwall region in the cases of the fully turbulent and transitional flow. The vortex flow pattern is complex enough – in this area we may observe 2 focuses, corresponding to the two counter-rotating vortices in the base region of the trailing edge (a 2D separation), and abound 7 pairs of saddle and spreading points, which indicate the position of the vortex zones of a 3D separation. We will not describe such a flow in excessive detail; however, we note the most important of its features. First, despite the fact that in the transitional flow case the cross-flow inside the endwall boundary layer starts upstream when compared with the fully turbulent flow case, it penetrates in spanwise direction approximately to the same distance in both flow cases. Perhaps this is due to the fact that in the transitional flow case the boundary layer at the blade suction side is thinner than that of the fully turbulent flow case. At the same time, its profile downstream of the cascade throat becomes turbulent and more filled. The mentioned phenomena prevent the further advance of the cross-flow within the boundary layer along the blade suction side in the spanwise direction.

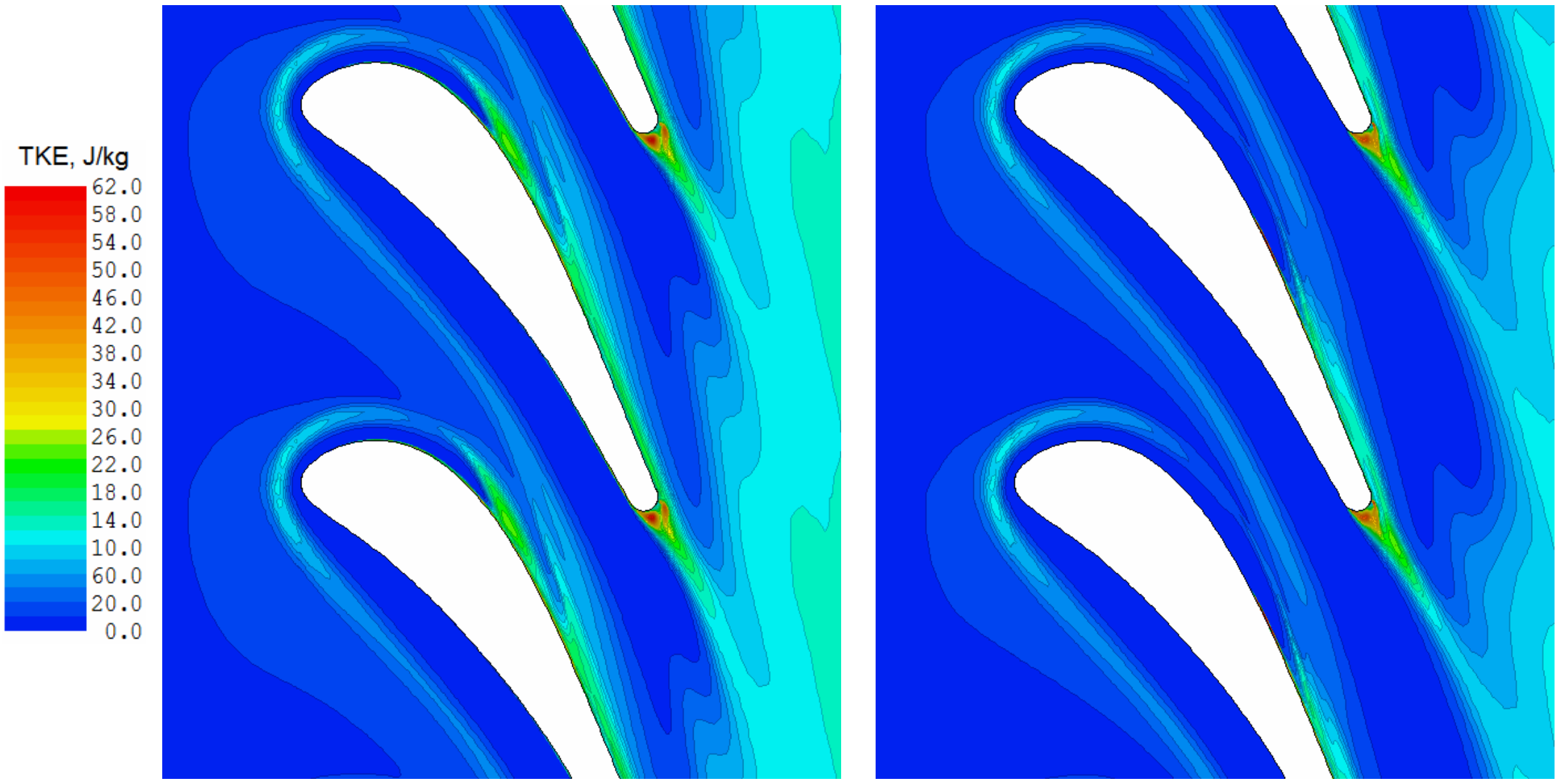

a b

Fig. 7. The TKE contours in the boundary layer
at the distance of 1 percent of the blade height from the endwall;
a – fully turbulent flow; b – transitional flow

Second, on the trailing edge, the spanwise cross-flows of alternating directions are observed, which are separated by singular points (two spreading points and one saddle point). This flow discontinuity results in a discreteness of the vortex wake downstream of the cascade. The cross-flow that is directed along the trailing edge to the mid-span sections of the blade channel forms an extensive discrete wake vortex, which we shall call the *main discrete wake vortex*. This cross-flow, and consequently, the main vortex are more intensive in the fully turbulent flow case. Closer to the endwall surface the *near-endwall vortex* of the opposite rotation direction is formed. As it is seen from the locations of the singular points at the trailing edge and endwall surfaces, the near-endwall vortex in the fully turbulent flow case, when compared with the transitional flow case, has a slightly larger dimension in the span direction, but it is smaller in the pitch direction. In the corner, which is formed by the trailing edge and endwall surfaces, the smaller *corner wake vortex* is induced.

Third, the typical dimensions of the corner vortex, which is induced by the passage vortex in the corner area between the suction side and the endwall surface, are larger in the case of the fully turbulent flow that is apparently also caused by a greater thickness of the boundary layer in this flow case.

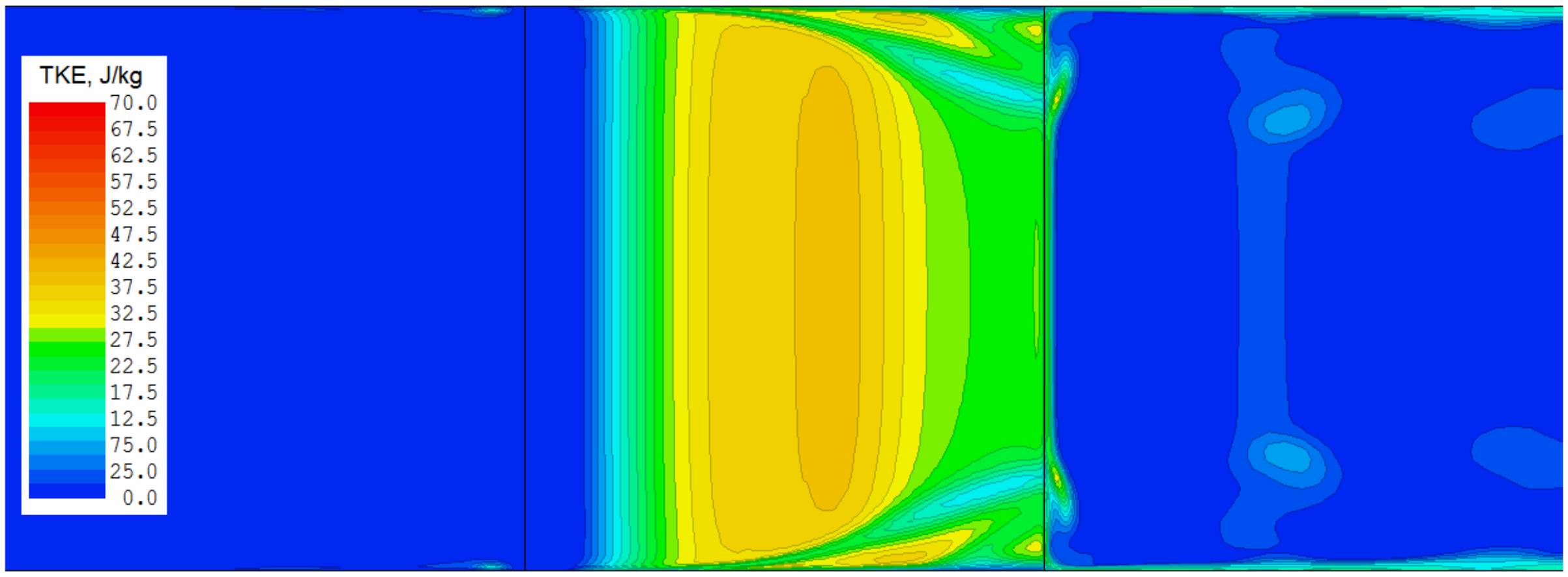

a

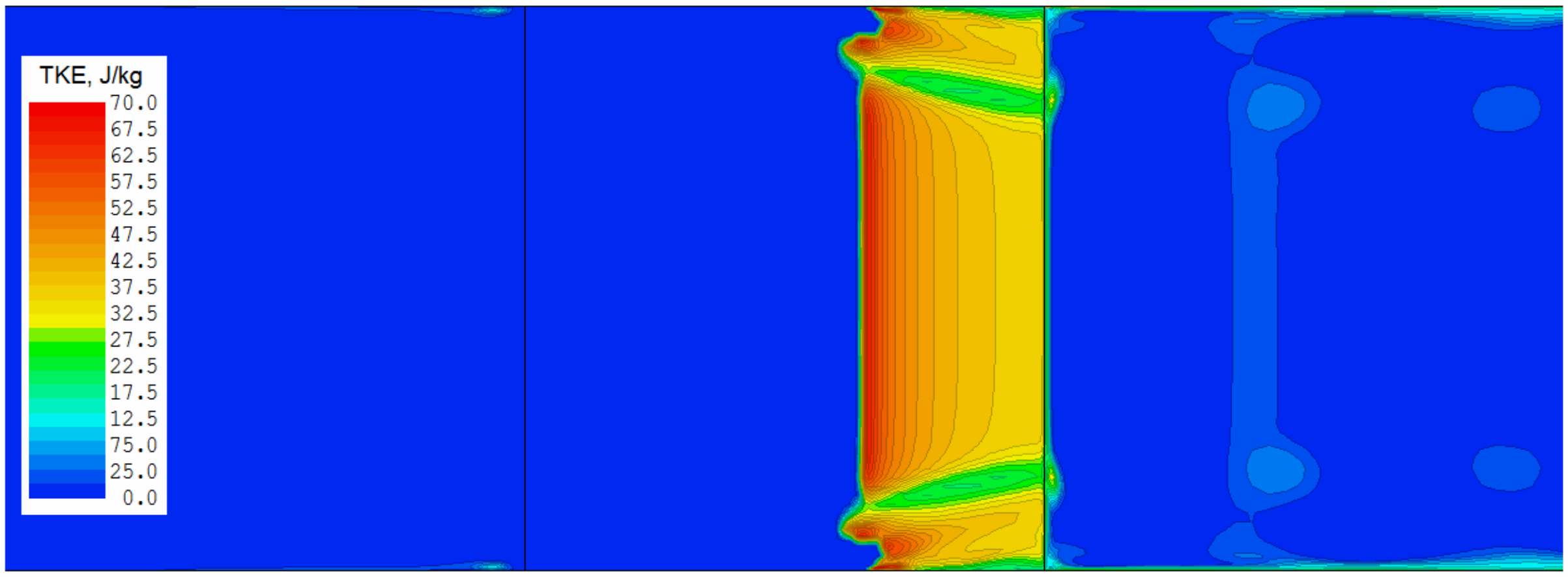

b

Fig. 8. The TKE contours at the distance of 0.1 percent of the cascade pitch from the blade suction surface
a – fully turbulent flow; b – transitional flow

Presented in Figure 10 are the contours of the entropy function $p/\rho^{\gamma}$ in a cross-section downstream of the trailing edge in fully turbulent and transitional flow cases. The flow is sufficiently symmetric relative to the middle of the cascade channel. The numbers in the Figure refer to the areas of increased entropy, the physical meaning of which will be explained below.

Demonstrated in Figure 11 are the secondary flow patterns in the cascade. The limiting streamlines at the endwall surface are shown to be similar to those in Figure 6. The entropy function contours at the perpendicular transversal surface are presented as it was done in Figure 10. Also in the endwall boundary layer region, the volume streamlines within the horseshoe, passage, and near-endwall vortices are displayed.

Analysis of these plots and of the entropy function distributions at several cross-sections in the proximity of the trailing edges and downstream of them allows us to note the following points.

The wake makes the main contribution to the growth of entropy, and, consequently, to the entropy losses in the considered cascade. The local maxima of the entropy function in Figures 10 and 11 correspond to the discrete wake vortices, described above. The effect of the secondary flows, as well as the dimensions of the regions occupied by them, are much smaller. This is explained by both a large thickness of the trailing edge and the low cascade loading.

The streamlines of the passage vortex, shown in Figure 11 in light color, come mainly in the high-entropy zone 1, which corresponds to the main discrete wake vortex. Both of these vortices have the same rotation direction and likely they are merged and mixed downstream of the trailing edge. It should be noted that in the transitional flow case, which is characterized by a larger intensity of the cross-flow in the endwall and suction side boundary layer, there is a greater dispersion of the passage vortex streamlines. As a result, in the fully turbulent flow case, the entropy values and the kinetic energy losses of the zone 1, which contains two vortices, are higher when compared with the transitional flow case, despite the fact that a cross-flow in the boundary layer and thereby the passage vortex are more intensive in the latter case.

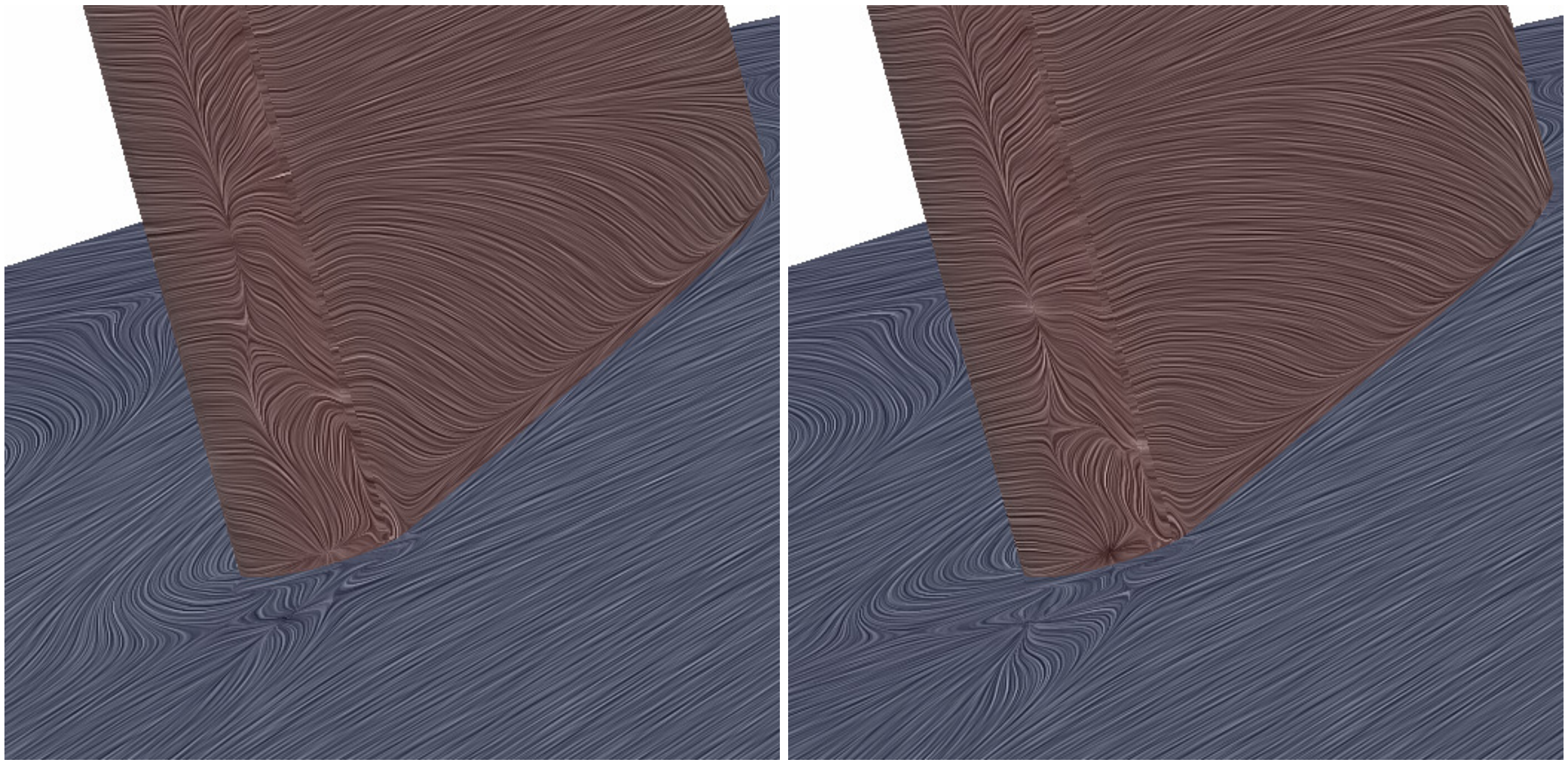

a b

Fig. 9. The limiting streamlines in the trailing edge/endwall corner region
a – fully turbulent flow; b – transitional flow

The streamlines that are painted in blue in Figure 11 extend from the near-endwall vortex downstream of the trailing edge into the high-entropy zone 2 of Figure 10. In this vortex, on the contrary, the streamlines are more dispersed in the case of the fully turbulent flow, whereas the intensity of the vortex is larger in the transitional flow case.

The high-entropy zone 3 in Figure 10 corresponds to the passage corner and wake corner vortices, which are not marked in Figure 11 because of their small size.

The branch of the horseshoe vortex (its core is indicated in red in Figure 11) that is formed near the blade pressure side extends towards to the lower boundary of the high-entropy zone 1. In the transitional flow case it is closer to the endwall surface when compared with the fully turbulent flow case. The branch of the horseshoe vortex that is formed near the blade suction side extends towards to the high-entropy zone 4, and in the fully turbulent flow case it is closer to the endwall surface.

In the case of the fully turbulent flow, the entropy values and the flow deviation angle in the mid-span sections of the wake are larger than those of the transitional flow case. This is due to the fact that in the fully turbulent flow case the boundary layers on both blade surfaces are thicker.

Figure 12 shows the distribution of the kinetic energy losses along the blade span. The red and blue lines indicate the present numerical results in fully turbulent and transitional flow cases, respectively. The positions of local maxima of the losses correspond to those of the entropy function distribution in the wake, shown in Figure 10. The values of the total and mid-span kinetic energy losses are given in the Table. The losses were determined at 40 percent and 100 percent of the axial chord downstream of the trailing edge. The latter corresponds to the exit boundary of the computational domain. It is seen that by taking into account the laminar-turbulent transition, we have improved the estimate of the total kinetic energy losses by more than 1 percentage point and of the mid-span kinetic energy losses by more than 0.5 percentage points.

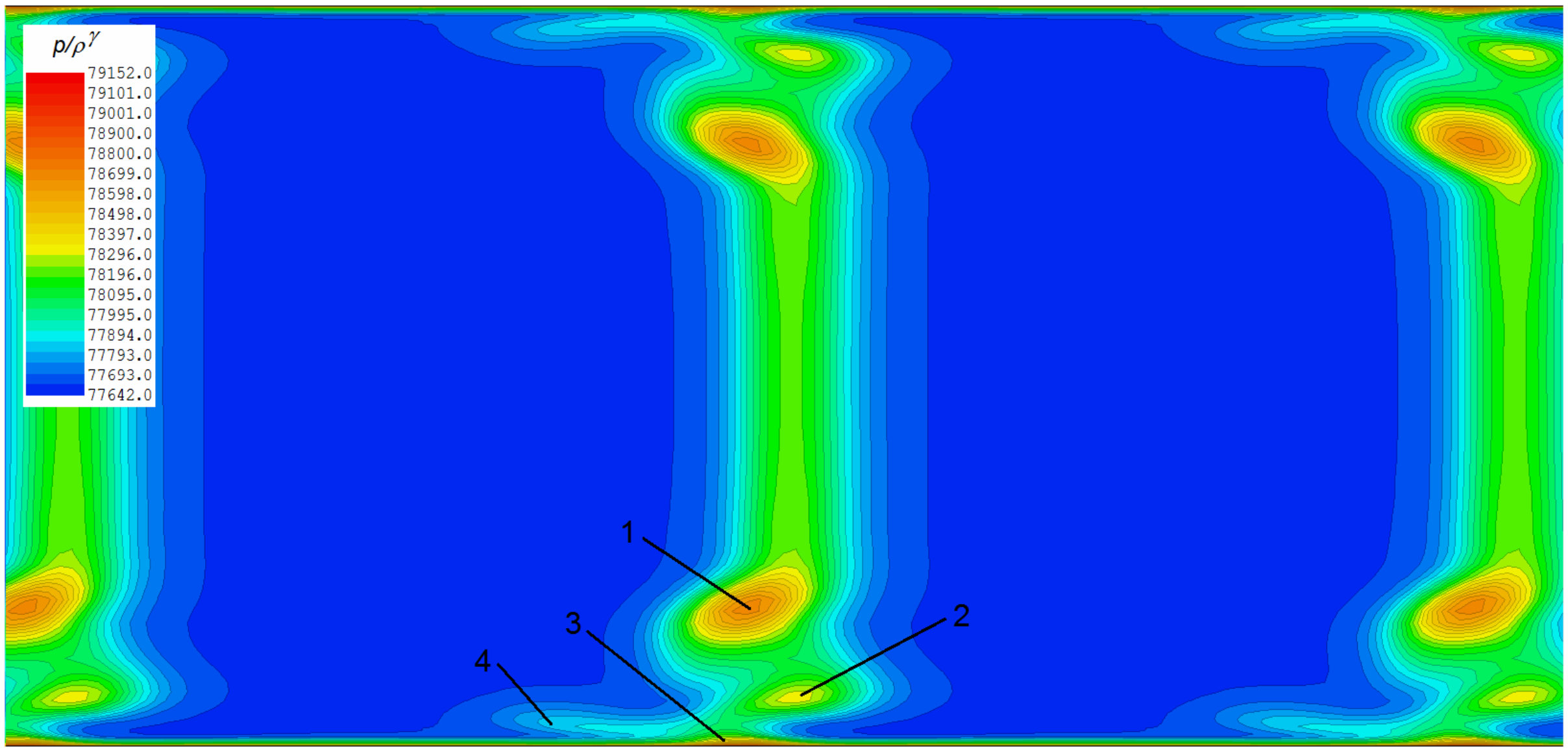

a

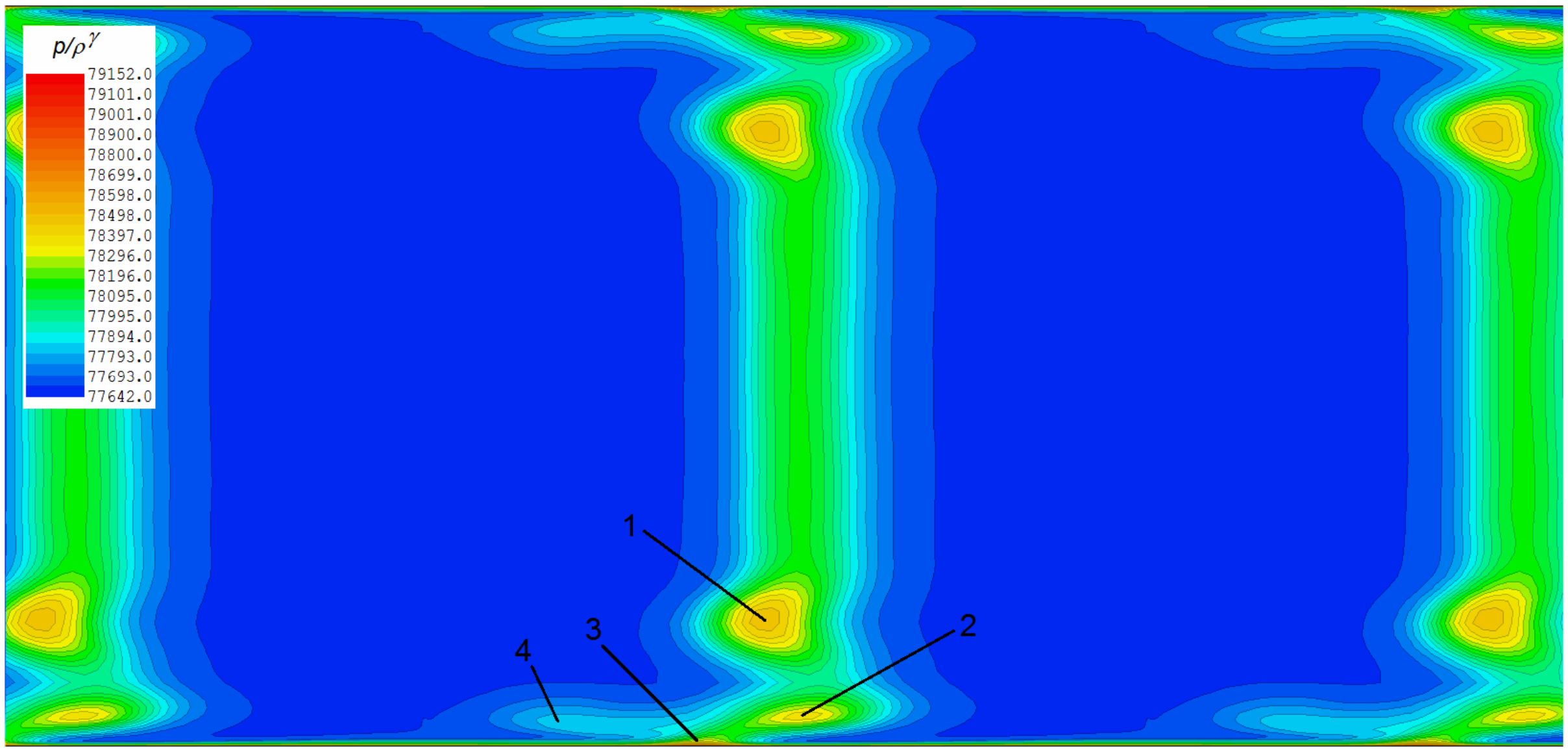

b

Fig. 10. The entropy function contours at the cross-flow section
at the distance of 20 percent of the axial chord downstream of the trailing edges;
a – fully turbulent flow; b – transitional flow

Table

The kinetic energy losses

| | Mid-span | | Total | |
|---|---|---|---|---|
| The distance downstream the trailing edges, the axial chord | 40 % | 100 % | 40 % | 100 % |
| Fully turbulent flow | 0.078 | 0.081 | 0.108 | 0.123 |
| Transitional flow | 0.070 | 0.073 | 0.093 | 0.109 |

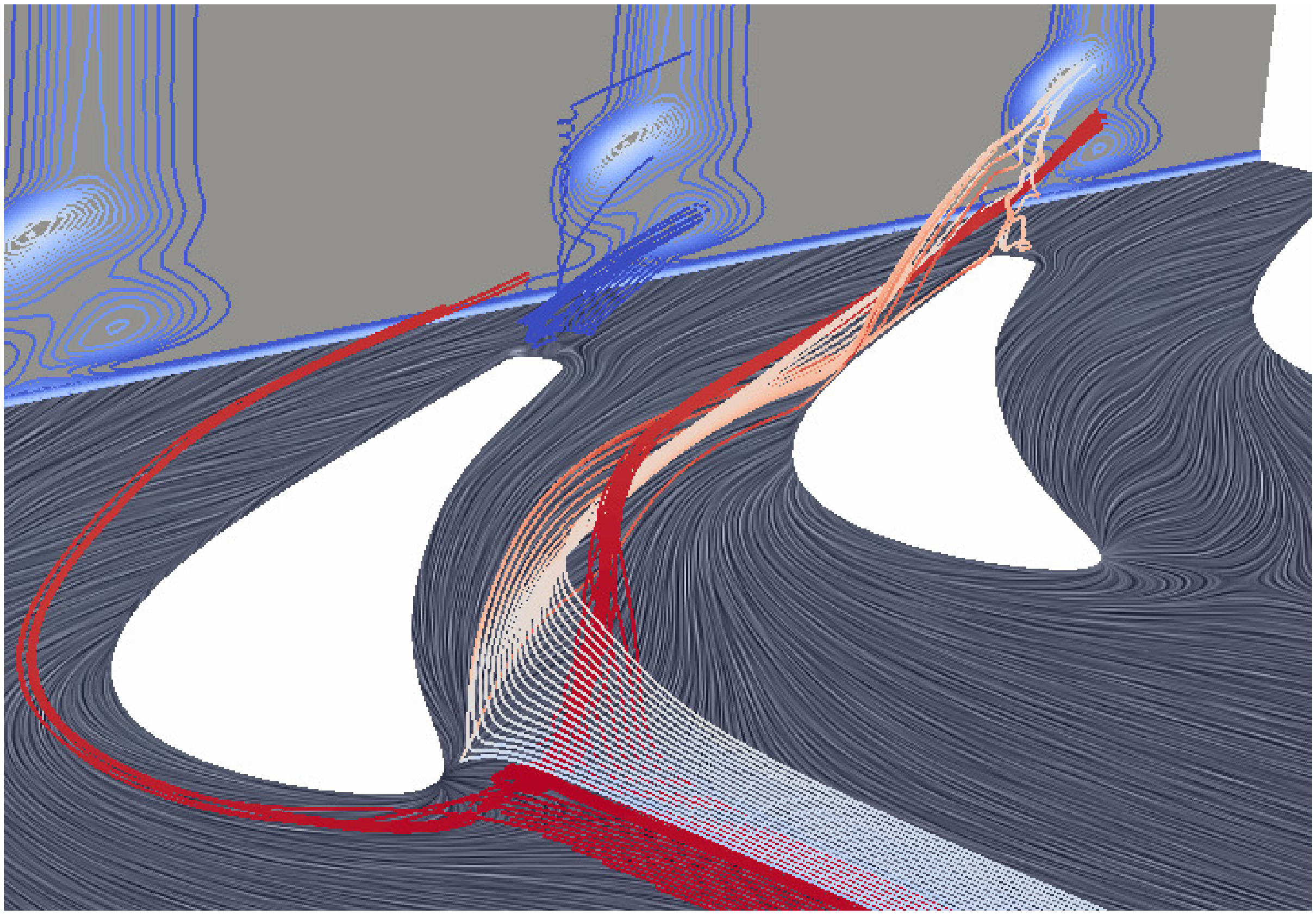

a

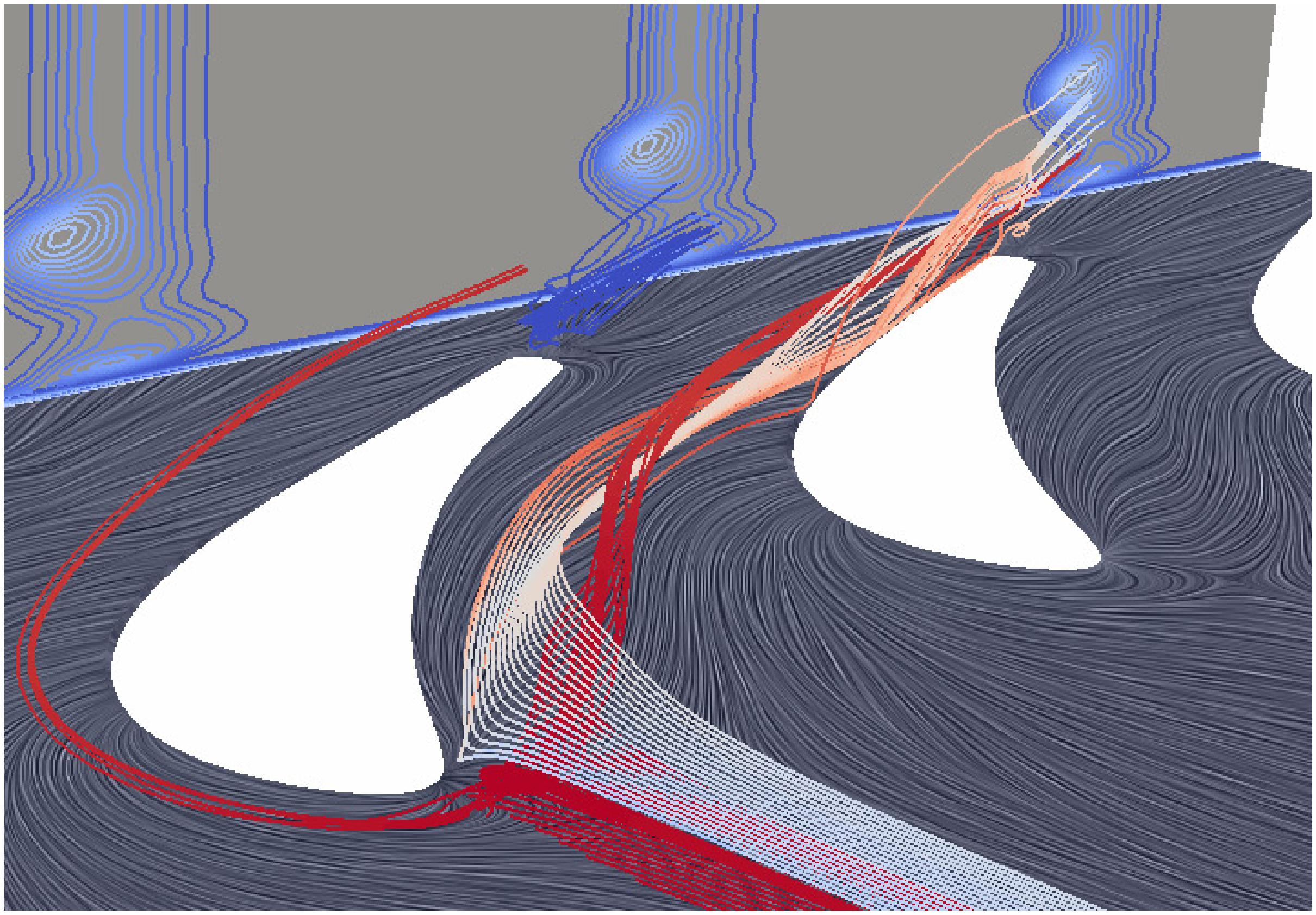

b

Fig. 11. The secondary flows structures
a – fully turbulent flow; b – transitional flow

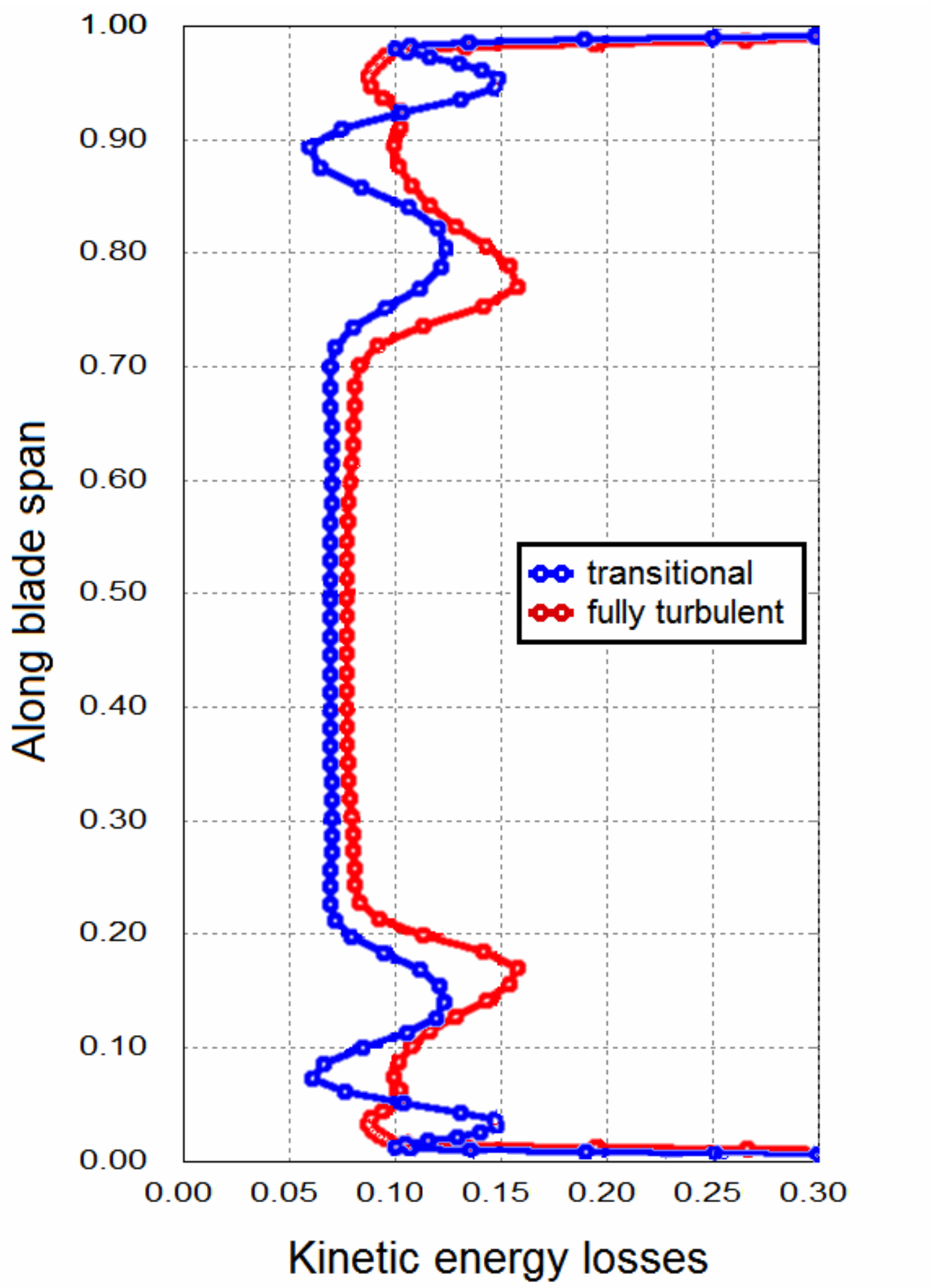

Fig. 12. The kinetic energy losses distribution along the blade span
at the distance of 40 percent of the axial chord downstream of the trailing edges

## 8. Conclusion

An algebraic transition PTM model permits the turbine cascade flow simulation taking into account the phenomenon of the laminar-turbulent transition. The location of the transition point, calculated using the approach suggested in this study is in a satisfactory agreement with the well-known experimental and numerical data. The TKE rise at both the suction and pressure blade sides is observed sufficiently downstream in the transitional flow case when compared with the fully turbulent flow case. In general, the physical representation of fully turbulent and transitional flows that is obtained numerically is consistent with the known ideas about this kind of flows.

Taking into account the laminar-turbulent transition, we have improved the estimate of the total kinetic energy losses by more than 1 percentage point and of the mid-span kinetic energy losses by more than 0.5 percentage points.

In the transitional flow case, due to the greater flow susceptibility to separation, the boundary layer cross-flow from the endwall surface to the blade suction side starts upstream, and the intensity of the near-endwall wake vortex is substantially higher, but, in general, the growth of the kinetic energy losses in the flow is significantly less. Consequently, the secondary flow patterns in the transitional and fully turbulent flow cases are somewhat different. Therefore, an important issue for the further research is the study of mechanism of the transition effect on the secondary flows and identifying various ways to reduce the kinetic energy losses in the turbine cascades by controlling the transition in the boundary layer.

The study found that transition modeling imposes severe restrictions on the adequacy of the turbulence model, the reliability and the operation speed of the numerical method, the resolution and the quality the computational mesh.

**Acknowledgement**

This work was performed under partial support of the Szewalski Institute of Fluid-Flow Machinery of the Polish Academy of Sciences, contract No 16/7/13/PS. The authors would like to thank Prof. P.Lampart for useful discussions.

**References**

[1] Hirsch C 2007 *Numerical Computation of Internal and External Flows: The Fundamentals of Computational Fluid Dynamics*, *2nd Edition*, Elsevier, Butterworth-Heinemann, 680 p.

[2] Schlichting H 1979 *Boundary-layer theory*, McGraw-Hill, New York, 817 p.

[3] Singer B A 1993 *Modeling the Transition Region, NASA Contractor Report,* 88 p.

[4] Elsner W 2007 *Transitional Modelling in Turbomachinery, J. Theor. and Appl. Mech.,* **45**(3), 539

[5] Sveningsson A 2006 *Transition Modelling - A Review, Technical report, Chalmers University of Technology, Gothenburg, Sweden,* 61 p.

[6] Langtry R B 2006 *A correlation-based transition model using local variables for unstructured parallelized CFD codes, Ph.D thesis,* University Stuttgart, 109 p.

[7] Menter F R 1994 *Two-Equation Eddy-Viscosity Turbulence Models for Engineering Applications, AIAA J.,* **32**(8), 1598

[8] Wilcox D C 1994 *Simulation of Transition with a Two-Equation Turbulence Model, AIAA J.,* **32**(2) 247

[9] Langtry R B and Sjolander S A 2002 *Prediction of transition for attached and separated shear layers in turbomachinery, AIAA Paper* 2002-3641, 13 p.

[10] Menter F, Ferreira J C, Esch T and Konno B 2003 *The SST turbulence model with improved wall treatment for heat transfer predictions in gas turbines, Proc. Int. Gas Turbine Congr.* IGTC2-3-TS-059, Tokyo, Japan, 7 p.

[11] Denissen N A, Yorden D A and Georgiadis N J 2008 *Implementation and Validation of a Laminar-to-Turbulent Transition Model in the Wind-US Code*, *NASA Technical Memorandum,* 215451, 36 p.

[12] Yershov S V 2008 *Realizability constraint for the SST k-ω turbulence model, Problems of Mechanical Engineering. Institute for Mechanical Engineering Problems of NAS of Ukraine,* **11**(2) 14 (in Russian).

[13] Yershov S V 1994 *Quasi-monotonous ENO scheme of high accuracy for Euler and Navier-Stokes equations, Matematicheskoye Modelirovaniye,* **6**(11) 63 (in Russian)

[14] Gryzun M N and Yershov S V 2013 *Numerical simulation of multi-dimensional compressible flows using the Newton's method. Power and Heat Engineering Processes and Equipment". National Technical University "Kharkov Polytechnic Institute" Bulletin,* (13) 38 (in Russian).

[15] Yershov S V 2015 *Free CFD code for turbomachinery,* http://sergiyyershov.com/ (accessed on 07/07/2015)

[16] Yershov S, Yakovlev V, Derevyanko A, Gryzun M and Kozyrets D 2012 *The development of new CFD solver for 3D turbomachinery flow computations. Cieplne Maszyny Przepływowe. Turbomachinery, Politechnika Łódzka,* Łódź, Poland (141) 15

[17] Ubaldi M, Zunino P, Campora U and Ghiglione A 1996 *Detailed Velocity and Turbulence Measurements of the Profile Boundary Layer in a Large Scale Turbine Cascade, International Gas Turbine and Aeroengine Congress and Exhibition, ASME* 96-GT-42, Birmingham, UK, 14 p.

[18] Malan P, Suluksna K and Juntasaro E 2009 *Calibrating the γ-Reθ Transition Model for Commercial CFD, AIAA Paper* 2009-1142, 20 p.

[19] Paraview, http://www.paraview.org/ (accessed on 07/07/2015)